# Casimir force: an alternative approach

P. R. Silva – Departamento de Física – ICEx – Universidade Federal de Minas Gerais
C. P. – 702 – 30123-970 – Belo Horizonte – MG – Brazil
e-mail 1: prsilvafis@terra.com.br – e-mail 2: prsilva@fisica.ufmg.br

ABSTRACT – The Casimir force between two parallel uncharged closely spaced metallic plates is evaluated in ways alternatives to those usually considered in the literature. In a first approximation we take in account the suppressed quantum numbers of a cubic box, representing a cavity which was cut in a metallic block. We combine these ideas with those of the MIT bag model of hadrons, but adapted to non-relativistic particles. In a second approximation we consider the particles occupying the energy levels of a Bohr atom, so that the Casimir force depends explicitly on the fine structure constant α. In both treatments, the mean energies which have explicit dependence on the particle mass and on the maximum occupied quantum number (related to the Fermi level of the system) at the beginning of the calculations, have these dependences mutually canceled at the end of them. Finally by comparing the averaged energies computed in both approximations, we are able to make an estimate of the value of the fine structure constant α.

## 1 – INTRODUCTION

In 1948, Casimir [1] showed that two parallel uncharged metallic plates separated by a close distance d suffer a force of attraction per unity of area given by

$$B = - (\pi^2 \hbar c)/(240 d^4), \qquad (1)$$

where ℏ is the reduced Planck constant and c the velocity of light in vacuum.

In his calculations, Casimir attributed this effect to the zero point fluctuations of the electromagnetic field, namely, the boundary conditions represented by the perfectly conducting plates suppresses some modes of the electromagnetic vacuum field which were present in the space region between the plates, before the set up of the metallic walls. In other words, Casimir was able to evaluate the difference in energy of the electromagnetic vacuum with and without the plates (which are both infinities), getting a finite energy which leads to a force per unit area given by (1).

An alternative interpretation of the Casimir force was proposed by Milonni and collaborators [2,3]. They considered that the virtual photons of the electromagnetic vacuum carry linear momentum. While the reflections of the photons of the zero point field inside the plates push them apart, the deflections of the outside modes pull them together. As there are more modes in the region outside the plates, the attractive character of the force prevails.

The experimental verification of the Casimir force was accomplished by Lamoreaux [4] in 1997, and recently by Bressi, Carungo, Onofrio and Ruoso [5]. Also, as pointed out by Lamoreaux [4], the Casimir force is closely related to the van der Walls attraction between dielectric bodies. Formally, Eq. (1) is obtained by letting the dielectric constant in the Lifshitz theory [6] approach infinity, which is an appropriate description for a conducting material. However the van der Walls force is always attractive, whereas the sign of the Casimir force is geometry dependent [7,8,9].

We judge worth to salient the the following considerations by Lamoreaux [10] in a Resource Letter of 1999 about the Casimir force: " The simplicity of Casimir derivation leads one to ascribe a certain reality to electromagnetic field zero-point fluctuations, implying that the Casimir force is an intrinsic property of space. However, there is a point of view that the attractive force is due only to the interactions of the material body themselves, as suggested by the Lifshitz derivation." It seems that the last point of view is shared by Jaffe, in a paper [11] where he states that: "Casimir effects can be formulated and Casimir forces can be computed without reference to zero point energies." We will comment about these different points of view latter on in this work

In this paper we intend to evaluate the Casimir force per unit of area between two parallel uncharged and closely separated metallic plates through alternative treatment to those usually considered in the literature.

In section 2, we consider the free electrons in metal filling the energy levels of fermionic particles in a cubic box and treat the free space between the plates as a suppression of this material media. In other words, we consider the cutting of a cavity in a metallic block. In the development of this task, the explicit information on the maximum quantum number (related to the energy of the Fermi level) and the mass of the free electrons which appear at the beginning of the calculations are mutually cancelled at the end.

In section 3, equations of motion are proposed as a means to justify the cancellations undertaken in section 2.

In section 4, the structure of the energy levels of the Bohr atom are used in a way similar to that done in section 2, where the energy levels of particles in a box were considered. The cancellations between the maximum quantum number and the mass of the free particles also take place there. The difference is that now the expression for the Casimir force displays an explicit dependence on the fine structure constant

By comparing the mean energies of the particles obtained in sections 2 and 4, we were able to estimate a value for the fine structure constant, α. This task is accomplished in section 5.

Finally we reserve section 6 for the concluding remarks.

2 – CASIMIR FORCE AND THE CUBIC BAG

Let us quote the following introductory words of the Chapter 6 of Kittel [12] on the free electron Fermi gas: "We can understand a number of important physical properties of metals , particularly the simple metals, in terms of the free electron model. According to this model the most weakly bound electrons of the atoms become the conduction electrons. Forces between conduction electrons and ion core are neglected in the free electron approximation; all calculations proceed as if the conduction electrons were free to move every where within the specimen. The total energy is all kinetic energy; the potential energy is neglected."

On the other hand in the MIT bag model [13,14], it is considered that free quarks are confined in the interior of the nucleon by the pressure of vacuum.

Inspired in the MIT bag model it is possible to imagine the free electrons in metal as a gas of non-relativistic particles confined by the vacuum pressure in the interior of a cubic box of edge equal to d.

Taking in account these considerations we can write the potential

$$V = N p^2/(2m) + B x^3, \qquad (2)$$

where the first term is the kinetic energy of *N* electrons of conduction, p its momentum, m its mass and B the pressure imposed by the vacuum on the bag. By using the uncertainty relation, $p = \hbar/x$, it is possible to write the potential (2) in the form

$$V(x) = A/x^2 + B\,x^3, \qquad (3)$$

where A is a constant which does not depend on x.

Taking the extremum of (3) with respect to x, we obtain

$$Bd^3 = (2/5)\,V(d) = (2/5)\,E_{av}, \qquad (4)$$

where we have identified V(d) with the averaged energy of the free electrons Fermi gas. Now for non-relativistic particles in a cubic box of edge d, their energy levels can be written as

$$E = [h^2/(8md^2)]\,(n_x^2 + n_y^2 + n_z^2), \qquad (5)$$

being $n_x$, $n_y$, and $n_z$ integer quantum numbers. By taking in account the cubic symmetry of the problem and the Pauli exclusion principle, we write the energy of the Fermi level as

$$E_F = [3h^2/(8md^2)]\,(n_x^2)_{max} = [3h^2/(32md^2)]\,N^2, \qquad (6)$$

where we have considered $(n_x)_{max} = N/2$, as the maximum occupancy number of the $n_x$ level.

In order to evaluate the mean energy, $E_{av}$, of the particles let us take

$$E(n) = [3h^2/(8md^2)]\,n^2 = E_1\,n^2, \qquad (7)$$

by putting $n_x = n_y = n_z = n$, in equation (4).

For $N/2 \gg 1$ (please see [15]), we can write

$$E_{av} = (1/N)\int_0^{N/2}(2E_1\,n^2 dn) = \tfrac{1}{3}\,E_F = [h^2/(32m)](N/d)^2. \qquad (8)$$

Comparing (8) and (4), we obtain

$$Bd^3 = [h^2/(80md^2)]\,N^2. \qquad (9)$$

Turning to the problem of determine the force per unit of area between two parallel uncharged plates, we may consider that if we cut in a metallic block a slice of thickness d, the suppressed free electron gas which occupied this slice are no more able to balance the vacuum pressure. Indeed if we cut a cubic cavity of edge d, the absence of the pressure due to the free electrons gas in the cavity can not balance the external vacuum pressure. In this way we propose that the attractive force per unit of area is given by B of relation (9), once the number N is properly computed.

We observe that in relation (9), B is a function of the electron rest mass m. As B is the pressure imposed by the quantum vacuum and therefore must not depend on massive fields we expect that the B dependence on m and N could be mutually canceled.

In reference [16] the de Broglie frequency was interpreted in terms of a harmonic oscillator applied to a material particle. The zero-point energy of this oscillator can be written as

$$E_0 = \tfrac{1}{2} m c^2 = p_0 c. \tag{10}$$

In the right side of (10) we have written this energy in terms of the energy of an equivalent photon of momentum $p_0$. We can also write

$$\lambda_0 = h/p_0 = 2h/mc. \tag{11}$$

Therefore the mass of the electron and the quantum number N can be eliminated from equation (9), by making the choice

$$N^2 = d/\lambda_0 = mcd/2h. \tag{12}$$

Inserting (12) into (9) we get

$$B = (3\pi\hbar c)/(240 d^4). \tag{13}$$

We observe that we have obtained in (13) the absolute value of the force per unit of area between the parallel uncharged metallic plates, but its character is attractive as seems to be implicit in the present analysis of the problem. Comparing (13) with (1) we see that we have, in the present treatment, obtained an approximate value for B which corresponds to 95 percent of the exact value calculated by Casimir [1].

3 – THE $N^2$ CHOICE: POSSIBLE FORMALIZATIONS

It would be worth to justify the choice of $N^2$ (see equation (12)), which led to the mutual cancellation of the mass term and the quantum number N appearing in the expression for B.

The free electrons of conduction obey the Fermi statistics and its relativistic treatment requires the use of the Dirac equation. However let us consider a scalar field $\Psi$ satisfying the equation of motion:

$$\partial\Psi/\partial x + c^{-1}\partial\Psi/\partial t = (mc/2h)\Psi - d^{-1}\Psi |\Psi^*\Psi|^2. \tag{14}$$

We observe that the linear terms of this equation (the first three terms) are similar to the Dirac equation, except by the absence of the spinorial character in the field $\Psi$.

A possible solution of (14) is

$$\Psi(x,t) = \sqrt{N}\, e^{i(kx + \omega t)}. \tag{15}$$

Inserting (15) in (14) and by setting $c = -\omega/k$, we obtain

$$|\Psi^*\Psi|^2 = N^2 = mcd/2h. \tag{16}$$

Therefore the result obtained through (14) is consistent with the choice made in (12).

Alternatively it is possible to think N as a variable satisfying a chemical-reaction-like equation, namely

$$v^{-1}\partial N/\partial t = mc/2h - d^{-1}N^2. \tag{17}$$

The stationary solution of (17), given by $\partial N/\partial t = 0$, leads again to

$$N^2 = mcd/2h, \tag{18}$$

which reproduces the results of (12) and (16). In the chemical-reaction equation given by (17), the $mc/2h$ term plays the role of a constant source field and v is a constant with a dimension of a velocity.

## 4- THE DEPENDENCE ON THE FINE STRUCTURE CONSTANT

As was pointed out by Jaffe [10], the Casmir force can be calculated without reference to vacuum fluctuations, and like all other observable effects in QED, it vanishes as the fine structure constant α goes to zero. It is the purpose of this section to pursue further on this subject.

One of the simplest models which exhibits energy levels dependence on the fine structure constant is the Bohr atom, namely

$$E_n = -\tfrac{1}{2}(\alpha^2 mc^2)/n^2 = -E_1/n^2. \tag{19}$$

Putting $n_{max} = N/2$, the Fermi energy of this structure reads

$$E_F = E(n = N/2) = -4E_1/N^2. \tag{20}$$

The mean energy could be estimated as

$$E_{av} = (2/N)\int_0^{N/2}(-E_1 n^{-2}dn) = (2/N)E_1[(2-N)/N]. \tag{21}$$

In the limit as $N/2 \gg 1$, we have

$$E_{av} = -2E_1/N. \tag{22}$$

Now let us estimate the attractive force per unit of area. We have

$$Bd^3 = -(2/5)(\alpha^2 mc^2)/N = (2/5)E_{av}. \tag{23}$$

By taking $p_0 = (1/2)\alpha mc$ and $\lambda_0 = h/p_0 = 2h/(\alpha mc)$, it is possible to make the choice

$$N = d/\lambda_0 = (\alpha mcd)/2h. \tag{24}$$

Inserting (24) into (23), we finally obtain

$$B = -[8/(5\pi)](\alpha\pi^2\hbar c)(1/d^4) \approx -(\pi^2\hbar c)/(269 d^4). \tag{25}$$

In the right hand side of (25), we took $\alpha = 1/137$, and we observe that the estimated value is 89 percent of the exact value calculated by Casimir. However result (25) is in agreement with the prediction made by Jaffe [11], that the force per unit of area must go to zero as the fine structure constant α vanishes.

In the case of the Bohr model is also possible to write an equation analogous to the equation (14), relative to the particles in a box model. We have

$$\partial\Psi/\partial x + c^{-1}\partial\Psi/\partial t = [(\alpha mc)/2h]\Psi - d^{-1}\Psi|\Psi^*\Psi|. \qquad (26)$$

Again a solution of equation (26) is

$$\Psi(x,t) = \sqrt{N}\, e^{i(kx+\omega t)}, \qquad (27)$$

where

$$|\Psi^*\Psi| = N = (\alpha mcd)/(2h), \qquad (28)$$

and

$$\omega/k = -c. \qquad (29)$$

5- ESTIMATION OF ALPHA

Casimir [17] speculated about the role played by the zero-point fluctuations of the electromagnetic field and its relationship to the fine structure constant.

In the present work, we have estimated the Casimir force in two alternatives ways. First we considered the energy levels of a particle in a cubic box and after we used the energy levels of the Bohr atom. Taking in account the equivalency between these two treatments let us make the requirement that the absolute values of the two estimated mean energies are equal. Then we have

$$(E_{av})_{box} = (hc)/(64d), \qquad (30)$$

where we have considered (8) and (12). And

$$|(E_{av})_{Bohr}| = (2\alpha hc)/d, \qquad (31)$$

where we have considered (19), (22), and (24).

Making the equality between (30) and (31), we finally obtain

$$\alpha = 1/128, \qquad (32)$$

which is not far from the known approximate value of 1/137, for the fine structure constant at low energies. Coincidentally, the running coupling constant of the QED calculated at the mass energy of the $M_w$, the intermediate vector boson of the weak interaction, is approximately given by [18]

$$\alpha(M_w) \approx 1/128. \qquad (33)$$

## 6- CONCLUDING REMARKS

There is a dispute about the proper interpretation of the Casimir force [2,10]. The first one ascribes a certain reality to the zero-point fluctuations of the electromagnetic field, implying that the Casimir force is an intrinsic property of the free space (vacuum). The second one attributes the attractive force only to the interactions of the material bodies themselves as in the Lifshitz treatment [6]. It seems that in the present work we were able to contemplate both perspectives.

Finally in evaluating the fine structure constant α in this work, perhaps a conciliation between the two points of view was achieved.